\documentclass[a4paper]{article}
\usepackage{epsf}
\begin{document}

\author{Andrea Baldassarri}
\title{Non-equilibrium Monte Carlo dynamics of the Sherrington-Kirkpatrick mean field spin glass model}
\date{July 23, 1996}
\maketitle

\begin{abstract}
We present a numerical study of the
non-equilibrium dynamics of the Sherrington-Kirkpatrick
model.  We analize the overlap
distribution between the configurations visited at the time
$t$ and in particular its scaling behaviour with the size of the
system. We find two different non-equilibrium dynamical regimes.
The first is a proper {\em Out of Equilibrium Regime}, 
that is the relevant regime for the dynamics of an infinite
system. The second is an {\em Intermediate Regime} that
separates the {\em Out of Equilibrium Regime} from
equilibrium.  After studying
the crossover beetween the two regimes, we focus on the
{\em Out of Equilibrium Regime} and we
reveal some of the geometrical features of phase space.
According to recent analitical 
work, we find that the
asymptotic out of equilibrium energy density is the
equilibrium one. The same happens to 
the staggered-magnetisation, suggesting there is a deep
geometrical similarity between equilibrium and
out of equilibrium configurations.  A procedure,
that we call {\em clonation}, shows that the dynamics follows
sort of canyons, as opposed to a system of
independent traps.
\end{abstract}

\section{Introduction}

	In the last years, the experimental, numerical and
theoretical interest on the non-equilibrium properties of
spin glasses \cite{mezparvir:sglass,binyou:sglass,fisher:sglass}
 has increased.
Experiments have shown a very rich behaviour in many
different materials (see for example 
\cite{str:aging,lun:aging,vinhamoci:aging,vinhamociboucug:aging}). 
Some particular procedures have
focused on measuring the main feature of the slow spin glass
dynamics: the aging phenomena. 
Simulations carried  out on
tridimentional spin glass models
\cite{rie:review}
show some of the characteristics (slow dynamics, aging) 
seen in experiments, but suffer of a lack 
of a microscopic description.
A recent analitical work
\cite{cugkur:sk} on the mean-field Sherrington-Kirkpatrick model
({\bf SK}) contribute to fill this gap, and confirm that 
spin-glasses represent a ground for developing  non-equilibrium
 statistical mechanics.

	In the present work we study the Monte Carlo
heath-bath dynamics (with Metropolis sequential updating
algorithm) of the SK model
\cite{shekir:sk},
defined by the Hamiltonian:
\begin{equation}
H_{J}[\sigma]=-\frac{1}{2}\sum_{i\neq j}J_{ij}
\sigma_i\sigma_j
\label{hamiltonian}
\end{equation}
The $J_{ij}$ are random Gaussian variables with variance
$1/\sqrt{N}$.

	We test some recent analitical predictions
\cite{cugkur:sk, balcugkurpar:sk},
and get insight into the geometrical
properties of the phase space interested by the dynamics. 
For this aim we use the technique of {\em dynamics clonation}
\cite{bal:thesis,cugdea:clones,barburmez:clones}
that though experimentally unimplementable, is easy to implement
 numerically and give us insight in the nature of the dynamics.

One of the main purposes of this paper is to study the range of 
validity of the analytical solution obtained in the thermodynamical
limit ($N \rightarrow \infty$) when the size of the system is finite ($N$ finite).
We find the ($N$,$t$) scalings to have a mean-field like
 {\em Out of Equilibrium regime}, an {\em Intermediate regime} and the 
final {\em Equilibrium regime}.

This paper is organized as follows:
In Section II we present the definitions of various quantities we 
study in this paper and we discuss different 
limiting procedures leading to 
equilibrium or out of equilibrium dynamics.
Section III is devoted to the numerical results.
Finally in Section IV we present our conclusions.

\section{General Considerations}

	Before reporting the results of the simulations,
let us observe that during a simulation the measured
quantity $O$ always depends on the size of the system (the
number of spins $N$) and on time (the number of Monte
Carlo sweeps preceeding the measuring time $t$). In the
following we shall show this dependency explicitly:
$O(N,t)$.  The equilibrium value is given by the
{\em Asymptotic Limit} ({\bf AL})
$lim_{t\rightarrow\infty} O(N,t) = O(N)$. 
The thermodynamical 
analytical calculations  give the equilibrium value for an
infinite system, that is the {\em Thermodynamical Limit}
({\bf TL}) of the equilibrium quantity:
$\lim_{N\rightarrow\infty} O(N) = O_{eq}$.  So, if we are
interested in the equilibrium properties of the system we
have to consider the following ordered limit procedure:
\[
\lim_{N\rightarrow\infty} \lim_{t\rightarrow\infty}
O(N,t) = \lim_{N\rightarrow\infty} O(N) = 
O_{eq}
\]

	Now let us consider the opposite order of the
limits. If the system undergoes a phase transition, the
{\bf TL} can break the ergodicity of the phase space. In
such a case, the {\bf AL} of the {\bf TL} may be different
from the equilibrium thermodynamical value:
\[
\lim_{t\rightarrow\infty} \lim_{N\rightarrow\infty} O(N,t) = 
\lim_{t\rightarrow\infty} O(t) =
O_{o.eq} \neq
O_{eq}
\]
	Some recent analytical works \cite{cugkur:pspin,cugkur:sk} 
deal with this
opposite ordered limit procedure to describe the
out of equilibrium properties of spin glasses. We call the
asymptotic dynamics of an infinite system, in the sense of
this order of limits, the non-equilibrium dynamics.

	In a real simulation, we cannot perform neither  {\bf
AL} nor the {\bf TL}. We consider systems of growing
size, and for each one we consider an equilibration time
after which the system is equilibrated.
	The correct criterion for choosing the
equilibration time is not so trivial, especially for
systems like spin-glasses, that evolve extremely slowly as shown
 clearly by experiments.
We can consider the spin-glass model
equilibrated when the Two-Times Quantities {\bf 2TQ} do not
depend explicitly upon both times, but only upon the
time difference (for brevity Homogeneous Time Dependence).
For example if we consider the Two-Times Auto-Correlation
Function
\begin{equation}
C_N(t,t')=
\overline{<1/N\sum_{i=1,N} \sigma_i(t)\sigma_i(t')>}
\label{auto-correlaz}
\end{equation}
we are sure that the system forgot the initial $t=0$
randomly chosen configuration only when
$C_N(t,t')=C_N(t-t')$. Hereafter we follow the standard motation, indicating 
the thermal average (different noise realization, but fixed $J_{ij}$) with 
$< \bullet >$, and the sample average (different chosen $J_{ij}$) with 
$\overline{\bullet}$.

Unfortunately, the measurements of {\bf 2TQ} are very computer time
consuming, because we have to keep track of all previous
configurations of the system. 
It would then be better use an equilibration criterion
involving only One-Time Quantities {\bf 1TQ}. 
However, in a glassy system, a {\bf 1TQ} may reach the
definitive asymptotic value, very near or equal to the
 equilibrium value, even  when the system is very far
from equilibrium and, for example, the Auto-Correlation
Function (\ref{auto-correlaz}) 
is far from being Homogeneous in times. So just checking
for the asymptotic value of a {\bf 1TQ} can lead to erroneous
results.

\section{Numerical Simulations}

\subsection{The Overlap as the geometrical Clock}

It is well known \cite{mezparvir:sglass} that the
SK model (\ref{hamiltonian})
undergoes a phase transition at $T_c=1$. The order
parameter is the Parisi $q(x)$ function that is related to
the overlap distribution function of the equilibrium states
$P(q)$:
$P(q)=(\frac{dq}{dx})^{-1}$ 
(For a recent numerical analysis, see \cite{picrit:sk}).
To monitor the order parameter, $q(x)$, we measure the
first non-zero momentum of the $P(q)$ overlap distribution,
that in the zero field case is:
\begin{equation}
q_2(N,t)=
\overline{<[(1/N)\sum_{i=1,N} \sigma_i(t)\tau_i(t)]^2>}
\label{q2}
\end{equation}
i.e. the second momentum. Here $\sigma_i(t)$ and
$\tau_i(t)$ denote the configuration at time $t$ of two
realization of the dynamics ($J_{ij}$ fixed), with independent randomly
chosen initial configurations and independent
noise realization. Hereafter we shall call them {\em
replicas}.
 	
	The quantity $q_2$ is used to determine
the equilibration of the SK model for example in
\cite{macyou:sk}. They use the
relation $q_2= 1 -2 T E$ \cite{bramor:sk} (where
$E=\lim_{N\rightarrow\infty}\lim_{t\rightarrow\infty}[\overline{<H>}/N]$ is the
equilibrium energy density) to determine the times $t<t_0$ to
skip before attempting the equilibrium measure.
On the contrary, in this work we are interested in the 
{\em evolution before the equilibration}.

\begin{figure}[] 
\epsfxsize=12cm
\epsfbox{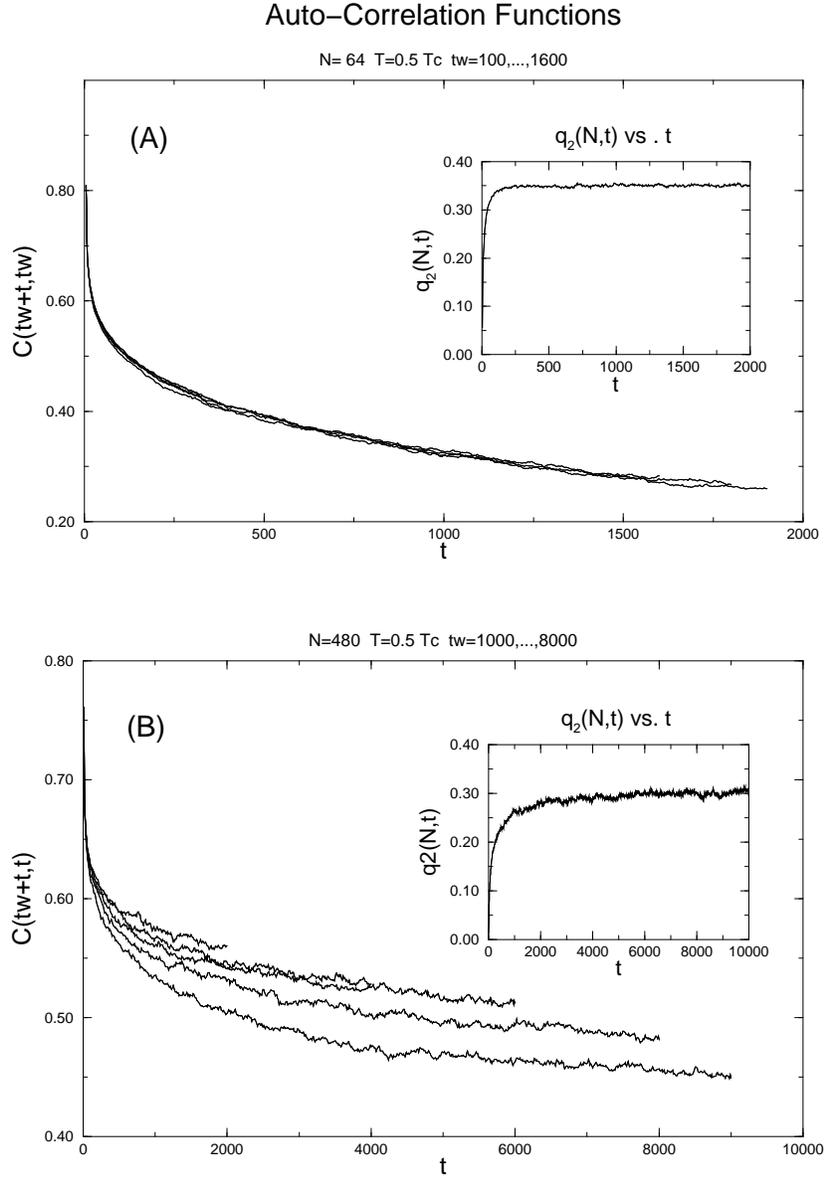}
\caption[]{The Auto-Correlation Functions (\ref{auto-correlaz}) become
time-homogeneous only when $q_2(N,t)$ (\ref{q2}) reaches its
equilibrium value. In (A) the system is equilibrated, in
(B) the system is out-of-equilibrium (the equilibrium value
of $q_2(N,t)$ is greater then $0.3$). The data refer to 200 
different samples (i.e. 200 different chosen $\{J_{i,j}\}$). 
For each sample we 
use at least 10 {\em replicas} (see text)
 to measure $q_2(N,t)$. \label{figorolo}}
\end{figure}

We check explicitly the hypothesis that  $q_2$ represents a
good ``clock'' for the dynamics, as shown in Fig.\ref{figorolo}. 
We concentrate much of our efforts studing this quantity
before it reaches the equilibrium value and in particular
its scaling behaviour respect to the size of the system.

	We simulate the dynamics of $q_2(N,t)$ for several
system sizes at fixed temperature. The value of
$q_2(N,t=0)$ is generally $1/N$, because the starting
configuration for each {\em replica} is choosen randomly and
independently. A different choice for the starting value of
$q_2$ is not determinant for the subsequent dynamics, apart
from  approximately the first hundred steps.

	The dynamics of $q_2(N,t)$ is characterized by a
monotonic growth towards the equilibrium value. We identify
two non-equilibrium regimes, characterized by different
scaling properties respect to the size of the system.
\begin{itemize}
\item[I)] {\em Out of Equilibrium Regime}.

	In the first regime,  $q_2(N,t)$ scales as $1/N$,
with the law
\[
q_2(N,t) \propto (1/N)\ln (t)
\]
	 In Fig.\ref{figovexn} we show the universal
quantity $N q_2(N,t)$ vs.  $t$ for systems of different
size. The temperature dependence is restricted to the
proportionality constant.
	We call this regime the {\em Out of Equilibrium
Regime}. In the {\em gedanken-simulation} of an infinite
system  $q_2(t)$ would be constantly zero.

\begin{figure}[] 
\epsfxsize=12cm
\epsfbox{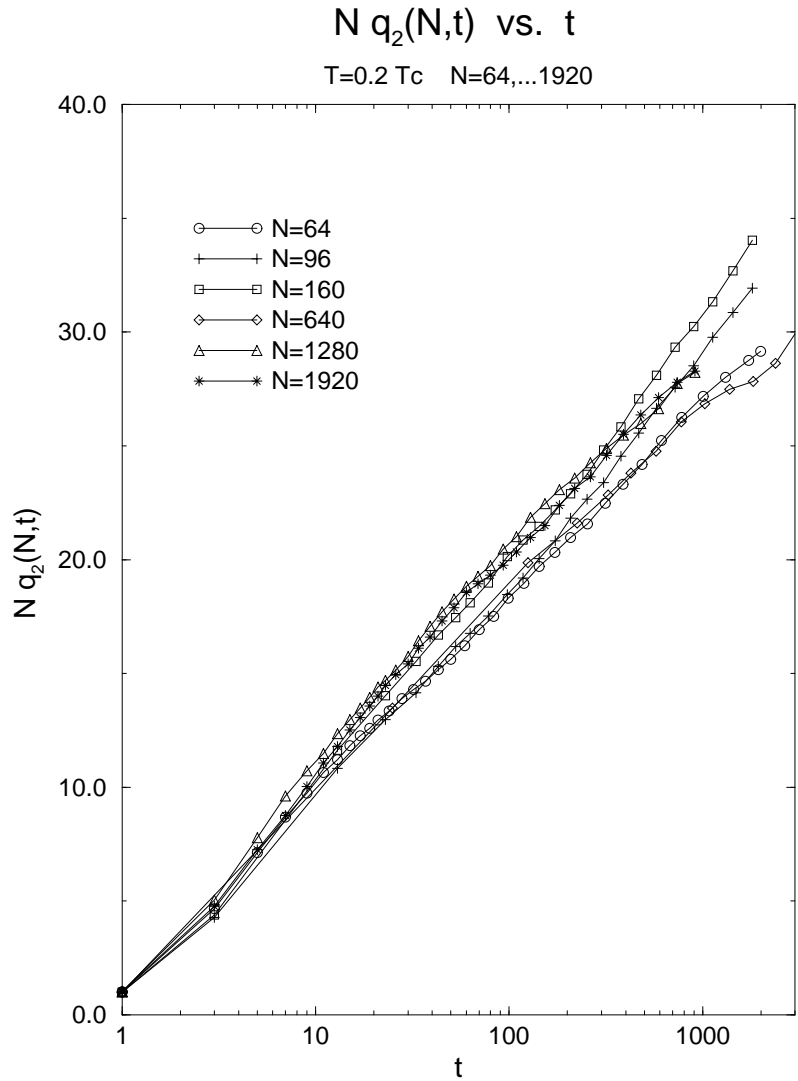}
\caption[]{The ``clock'' $q_2(N,t)$ in the {\em Out of Equilibrium
Regime} multiplicated by $N$ for systems of different size.
The data refer to a minimum of 10 replicas for each of 200
samples.
\label{figovexn}}
\end{figure}

\item[II)] {\em Intermediate Regime}.

	As the $q_2(N,t)$ reaches a sensible fraction of
its equilibrium value, the scaling behaviour changes.
We tried a fit for this scaling. For a fixed
temperature, we plot the quantity
$q_2(N,t)/q_2^{eq}(N)$ vs. 
$\ln (t)/N^{\alpha}$.  Varying $\alpha$, we looked
for the collapse of the curves associated to systems of
different size. In Fig.\ref{fignalfa} is shown
the result for
$T=0.4T_c$, that indicates a value $\alpha \approx 0.5$. 
But the value of $\alpha$ is temperature dependent.  For
different temperatures we found values in the range between
$1$ and $1/3$. We called this regime the {\em Intermediate
Regime}.
 
\begin{figure}[] 
\epsfxsize=12cm
\epsfbox{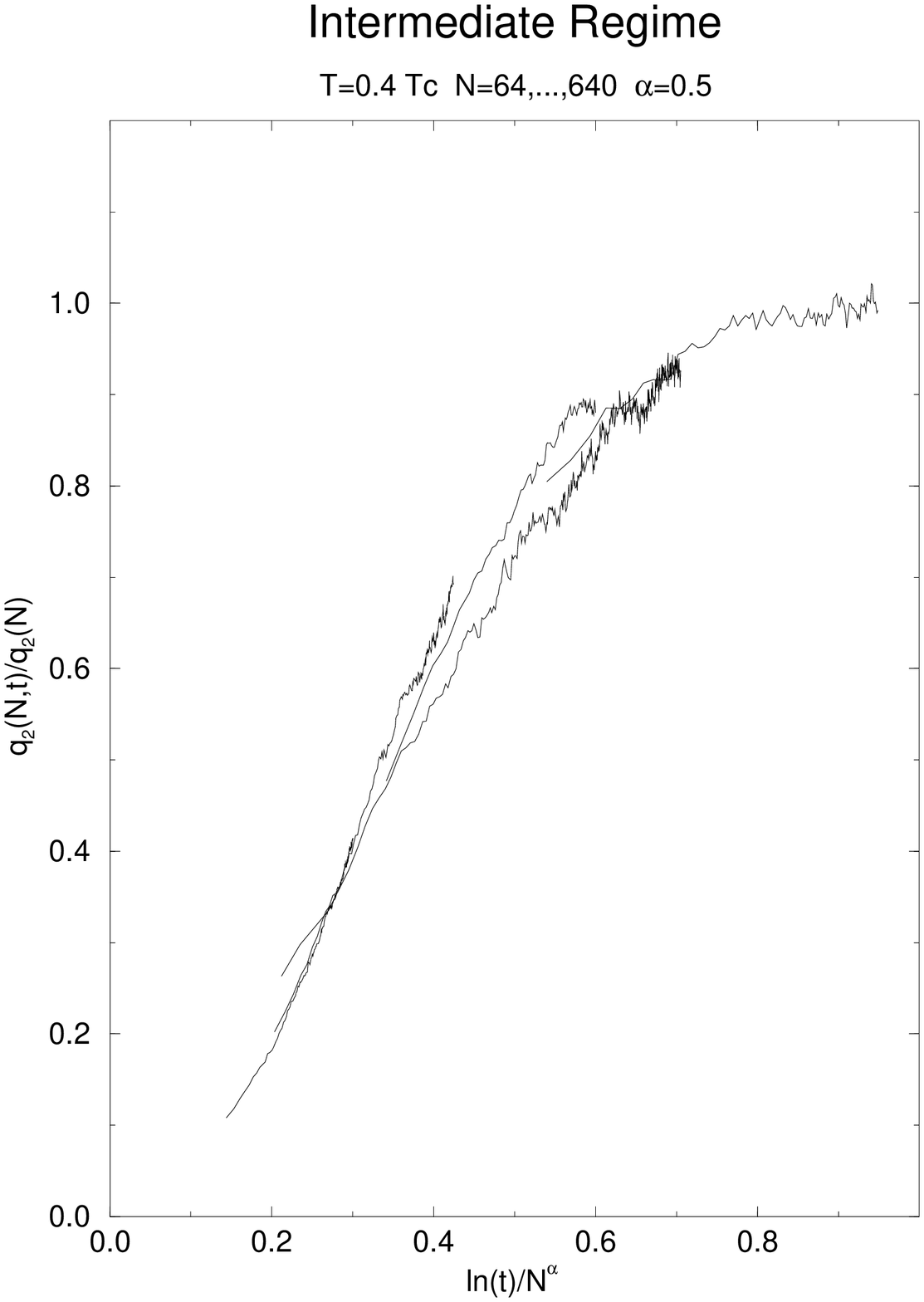}
\caption[]{In the {\em Intermediate Regime} $q_2(N,t)$
scales as $1/N^{\alpha}$. Here we plot $q_2(N,t)$ vs.
$\ln(t)/N^{\alpha}$ for the value $\alpha=0.5$, the more
appropriate for this temperature ($T=0.4T_c$).
\label{fignalfa}}
\end{figure}

\end{itemize}

\begin{figure}[] 
\epsfxsize=12cm
\epsfbox{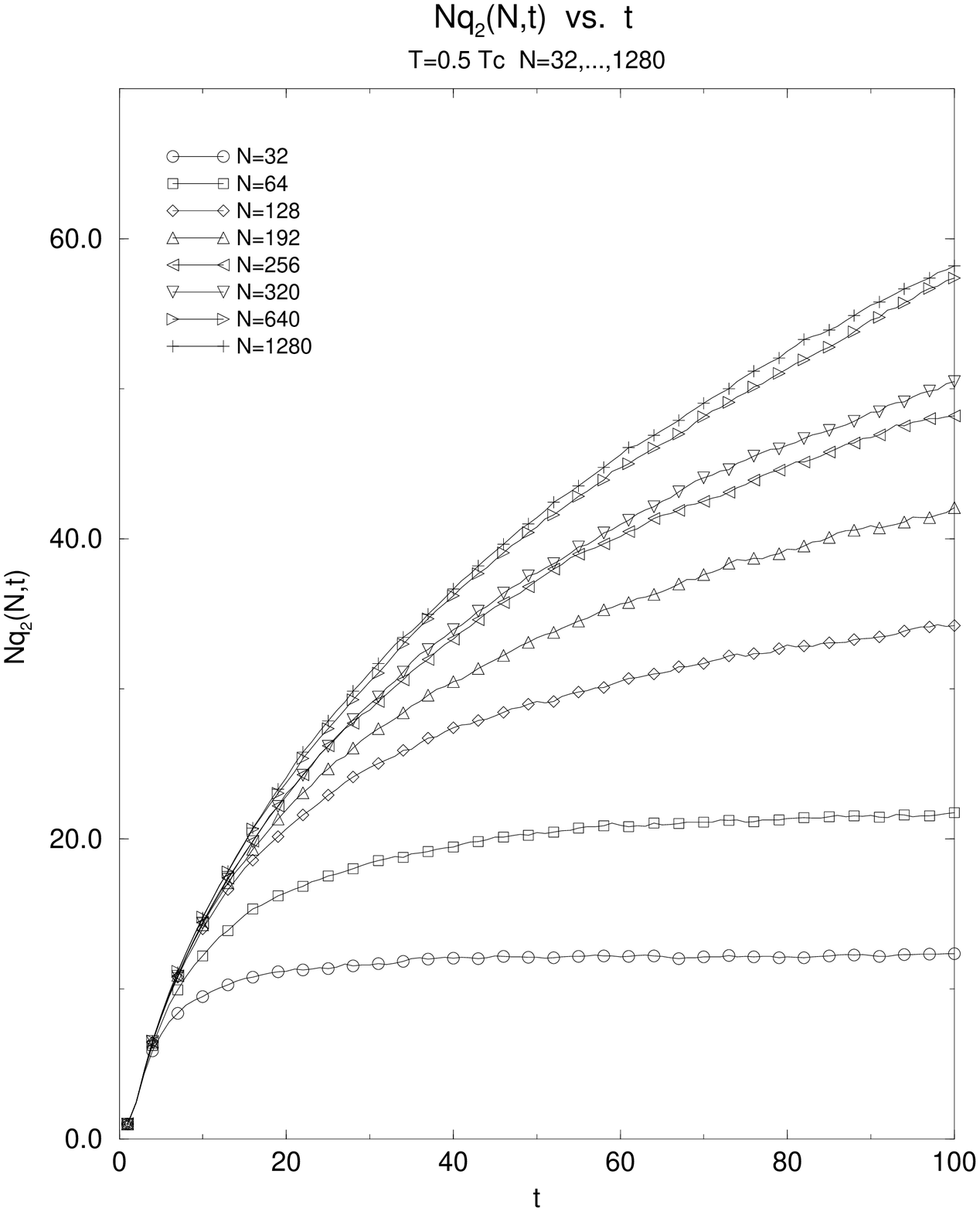}
\caption[]{Cross-over from the {\em Out of Equilibrium regime} to
the {\em Intermediate regime}. The plot represents the quantity $N
q_2(N,t)$ vs $t$ for several systems of growing size. Note
that the curves assciated to $N=640$ and $N=1280$ collapse,
indicating that the two systems are still both in the
{\em Out of Equilibrium Regime} for the whole time window. The curve
for $N=1280$ has been used as the limiting curve for the
determination of the Cross-Over time.
\label{figcross1}}
\end{figure}
\begin{figure}[] 
\epsfxsize=12cm
\epsfbox{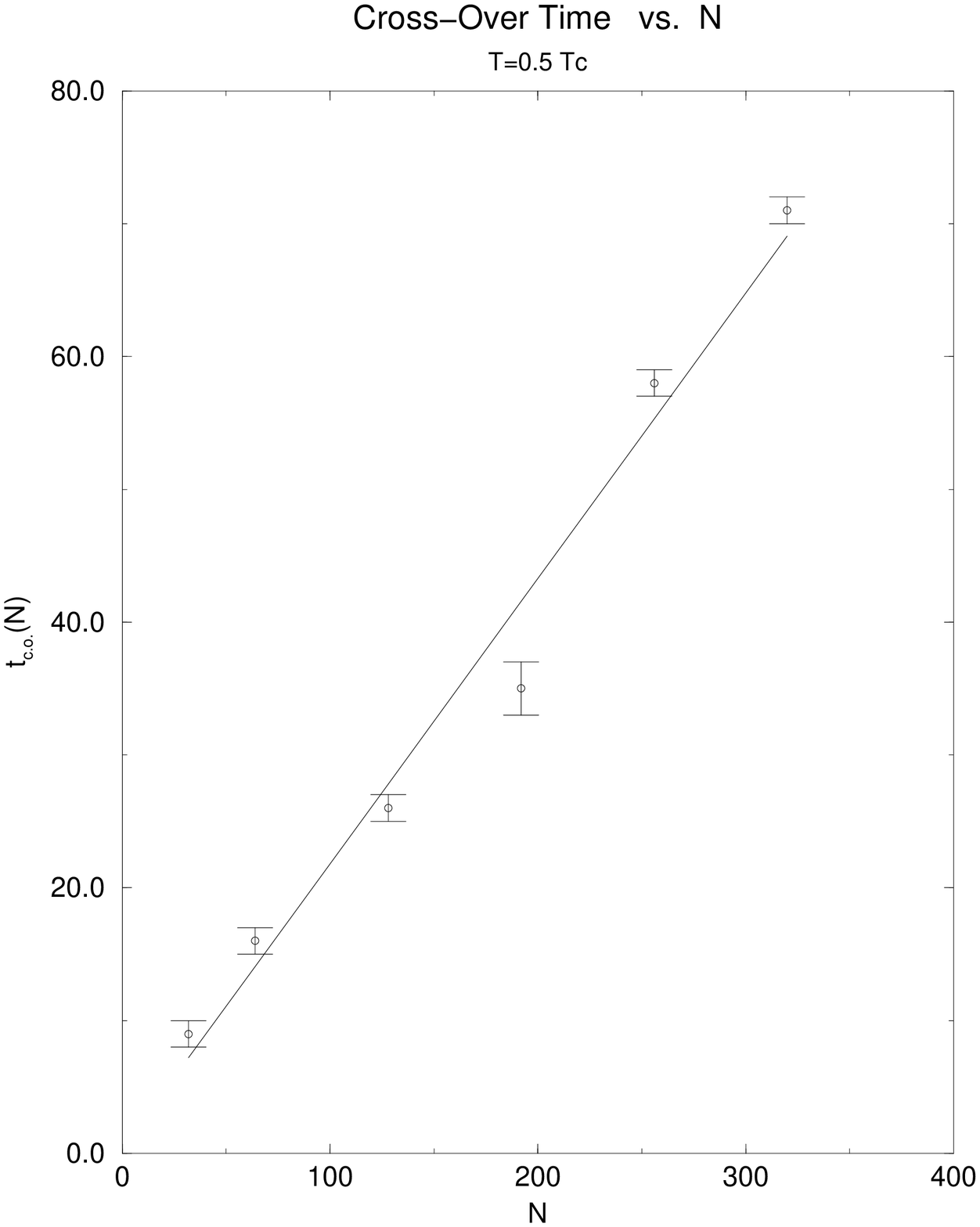}
\caption[]{Cross-Over time between the two non-equilibrium
regimes vs the size of the system $N$. The fit suggests a
linear behaviour.
\label{figcross2}}
\end{figure}

The cross-over between the two regimes is shown in
Fig.\ref{figcross1}. We plot the quantity $N q_2(N,t)$ that follows an
universal curve (at fixed temperature) in the first regime.  Larger
systems depart from this curve later, since the cross-over time grows
with the system size. We define the cross-over time as the time before
the difference between the $Nq_2(N,t)$ and the universal curve
$\lim_{N\rightarrow \infty} Nq_2(N,t)$ reaches a fixed tollerance
value. In Fig.\ref{figcross2} we show the cross-over time vs. the
size of the system. We guess a linear behaviour.

\begin{figure}[] 
\epsfxsize=12cm
\epsfbox{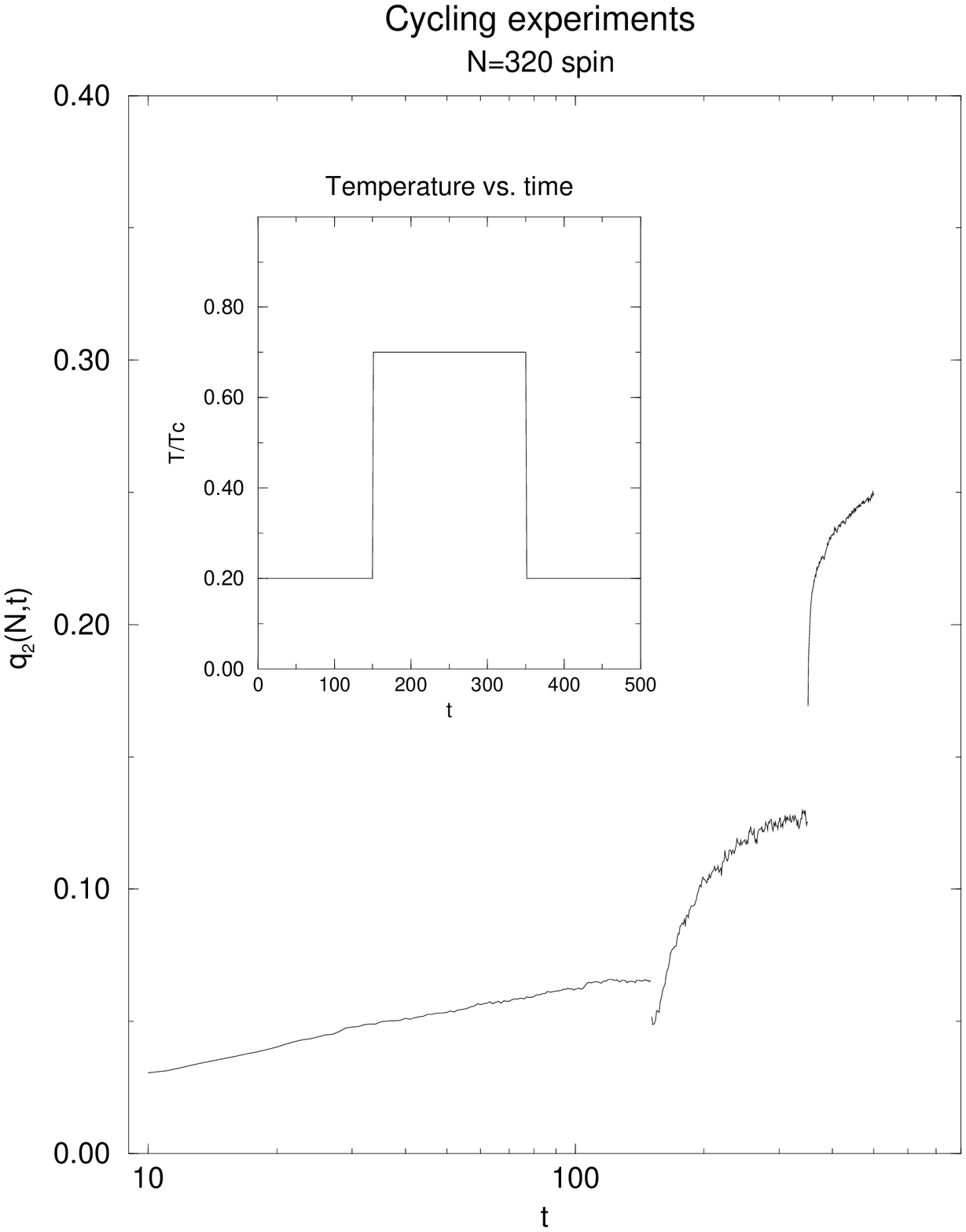}
\caption[]{Response of the ``clock'' of the dynamics
$q_2(N,t)$ to a cycle of temperature. In the inset we show
the temperature vs the time $t$. We have performed a short
heat pulse.
\label{figciclo}}
\end{figure}

In Fig.\ref{figciclo} we show the response of the
dynamical ``clock'' $q_2(N,T)$ to a ``positive'' cycle of temperature.
We rise the temperature to a still subcritical value for a
period, after that we restore it to the initial value. The
overall effect of the cycling procedure is that the system
goes closer to the equilibrium, since the $q_2(N,t)$
``clock'' 
reaches a higher value than the one obtained, at the same time, 
in the absence of the temperature cycle.

Such behaviour seems
different from the experimental results reported by
Hamman et al. in \cite{vinhamoci:aging}, where they see, for positive 
temperature cycles,
a restart of the dynamics, i.e. a loose of memory of the evolution 
before the heat pulse.  

However let us note that the response of the system
to the positive increase of the temperature is very different from
the response to the decrease.

\subsection{Out of Equilibrium Regime}

	As we are interested in the out-of-equilibrium
dynamics of the model, in the sense of the ordered  limit
procedure {\bf AL} after {\bf TL}, we study the dynamics
of some quantities in the first dynamical regime, i.e.  the
Out of Equilibrium Regime.


	The Energy density of the system at time $t$ is:
\begin{equation}
E(N,t)=(1/N)\overline{<H(t)>}
\label{energydensity}
\end{equation}	
	The relaxational dynamics of the Energy density for
the SK model has been extensively studied (for example in
\cite{kin:sk}). 
Here we want to focus into 
its asymptotic value. To this aim, we
reparametrize the energy density curve with respect to the
dynamical ``clock''
$q_2(N,t)$, that indicates the equilibration level of the
system at time $t$.

\begin{figure}[] 
\epsfxsize=12cm
\epsfbox{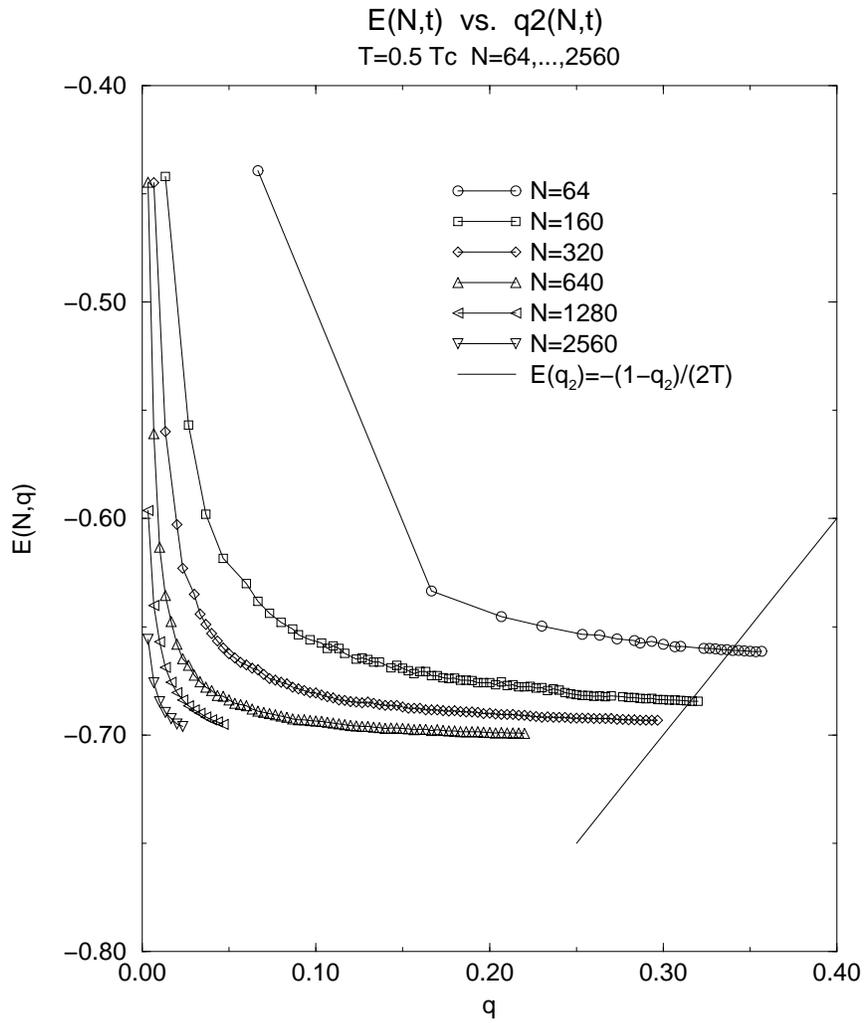}
\caption[]{Energy density relaxation curve in function of
the corresponding value of the ``clock'', i.e. $E(N,t)$ vs
$q_2(N,t)$.
\label{figeneove}}
\end{figure}

	The Figure \ref{figeneove} shows the result of such
reparametrization. The line is the equilibrium relation
$E=-\beta/2(1-q_2)$.  The first two curves refer to systems
that reach equilibrium at the end of the simulation. 
The others, with growing size, are farther from
equilibrium, but closer to the energy density equilibrium
value. 
A plausible proposal for the limiting curve associated to
an infinite system is a step function that, starting
from $E=0$, $q_2=0$, abruptely collapses to the constant
value $E=E_{eq}$ for $q_2>0$.
Besides it means that this {\bf 1TQ} does not
represent a good choice for the determination of the
equilibration time.

Our results disagree with those by Scharnagl et al. in 
\cite{opp:sk}. 
The results of their power law fit 
on the first 130 steps for the energy dynamics
of an infinite system (they use an interesting new procedure 
to simulate the dynamics of infinite sized systems \cite{eisopp:sk})
indicate an asymptotic energy value $e_{\infty}$ different from the 
equilibrium value for temperature below $0.5T_c$.

On the contrary, 
we find that the system relaxes to its equilibrium energy
density.
To corroborate our result, we summarize here 
previously published results \cite{balcugkurpar:sk}.

	Consider $\sigma_{\lambda}$ the projection of the
configuration vector
$\overrightarrow{\sigma}(t)=\{\sigma_i(t)\}$ on the set of
eigen-vectors $\overrightarrow{J}(\lambda)$ of the coupling
matrix $\{J_{ij}\}$.
We measure the so called staggered magnetisation, i.e.
\[
g_{N,t}(\lambda)=\overline{<(\sigma_{\lambda}(t))^2>}
\]
where $\sigma_{\lambda}(t)=\sum_{i=1,N} \lambda_i
\sigma_i(t)$,
$\{\lambda_{i}\}$ being the coordinates of the
$\hat{\lambda}$-eigenvector of the coupling matrix
($\sum_{j=1,N} J_{i,j}
\lambda_j = \hat{\lambda} \lambda_i$).   	

The staggered magnetization is a distribution that contains 
a lot of informations.
In particular its first momentum is the energy density (\ref{energydensity}),
 object of our previous discussion. More precisely:
\[
\lim_{N\rightarrow \infty} 
E(N,t)= - \lim_{N\rightarrow \infty} \int_{-2}^{2} 
\lambda \rho(\lambda) g_{N,t}(\lambda) d\lambda
\]
where $\rho(\lambda)$ is the  well known 
semicircle distribution for the eigenvalue of the $J_{ij}$
random matrix \cite{meh:matrix}.

\begin{figure}[] 
\epsfxsize=12cm
\epsfbox{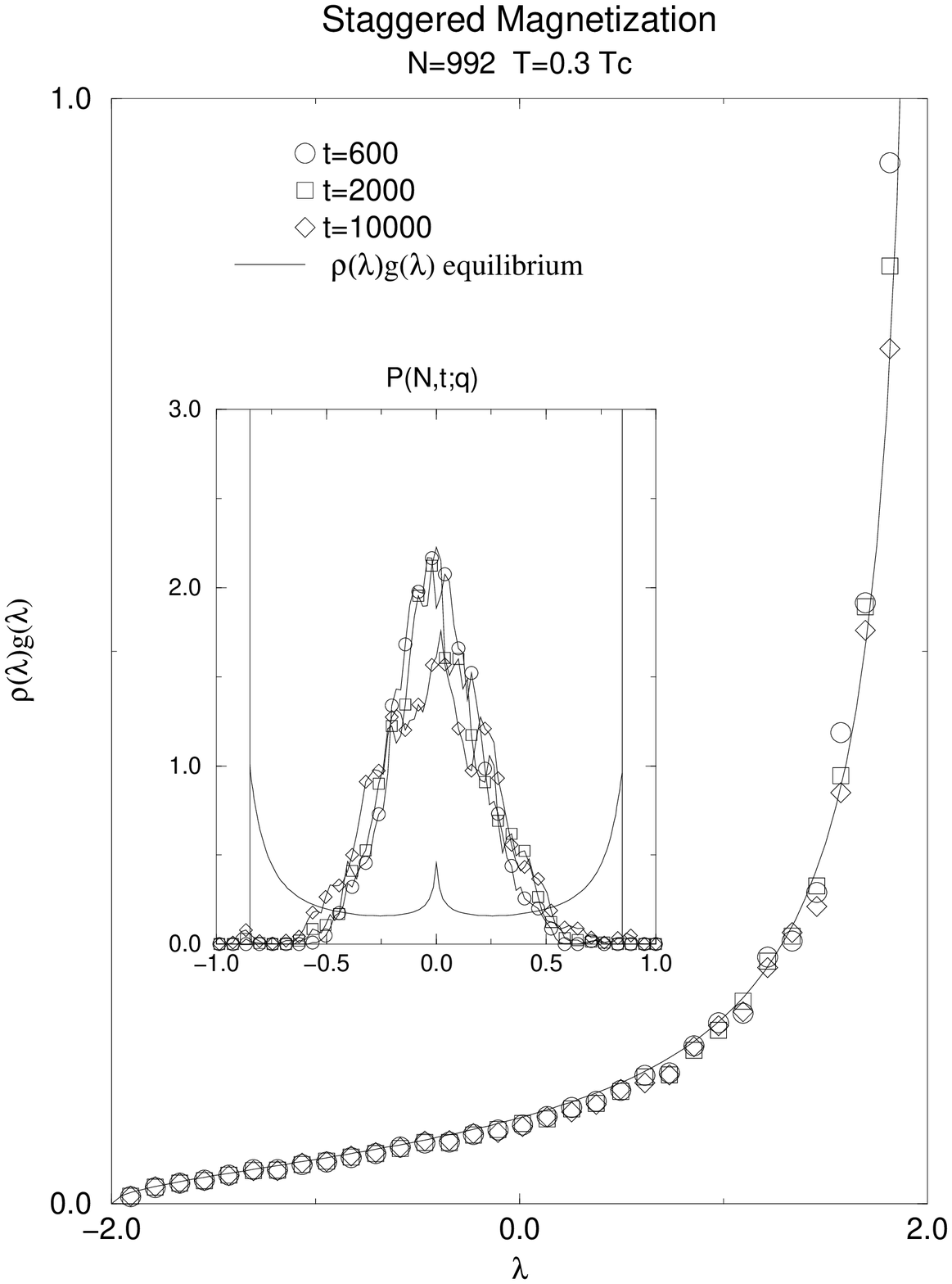}
\caption[]{Staggered magnetisation calculated at different
non-equilibrium times (from ref. \cite{balcugkurpar:sk}).
 In the inset we show the overlap
distribution for the same times. The full lines correspond
to the prediction of the replica ansatz for the
equilibrium.

\label{figstag}}
\end{figure}

In Fig.\ref{figstag} we show the
staggered-magnetization, multiplicated by the  
eigenvalue semicircle distribution $\rho(\lambda)$, compared with 
the equilibrium curve \cite{mezpar:self}. 
We note a very fast convergence of the
out of equilibrium staggered magnetization to the
equilibrium curve,
when the system is still very far from
equilibrium (as revealed by the $P(q)$ shown in the figure inset). 

\begin{figure}[] 
\epsfxsize=12cm
\epsfbox{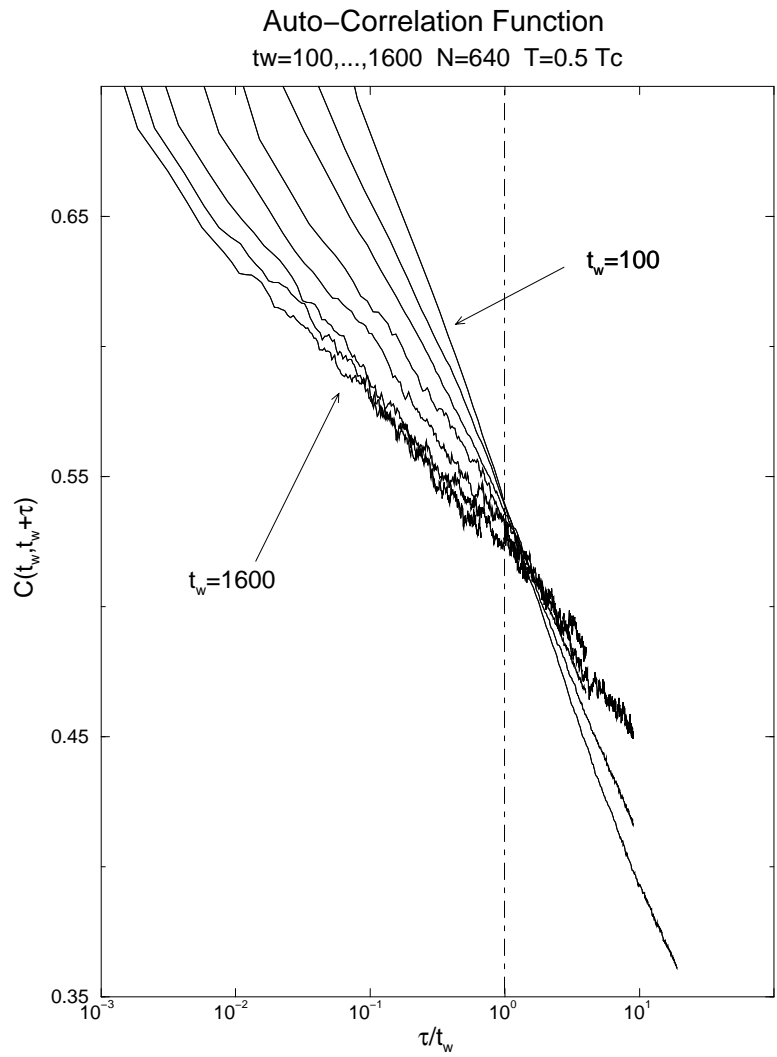}
\caption[]{Auto-correlation functions $C(t_w,t_w+\tau)$ vs $\tau/t_w$
for different $t_w$. We do not note a simple $\tau/t_w$
behaviour.
\label{figcor}}
\end{figure}

	The Auto-Correlation function 
(\ref{auto-correlaz}) 
shows aging behaviour (see Fig.\ref{figcor}), confirming the weak ergodicity
breaking hypothesis (\cite{bou:aging},\cite{cugkur:sk}):

\begin{eqnarray}
\nonumber
\lim_{\tau\rightarrow\infty}C(t_w,t_w+\tau)&=0&\\
\nonumber
\frac{\partial C(t_w,t_w+\tau)}{\partial \tau}&\leq 0
&\forall\textnormal{(fixed)} t_w \\
\nonumber 
\frac{\partial C(t_w,t)}{\partial t_w}&\geq 0
&\forall\textnormal{(fixed)} t>t_w 
\end{eqnarray}

Consider the fluctuations of the Auto-Correlation Functions:
\begin{equation}
F_N(t,t')=
\overline{<(1/N\sum_{i=1,N} \sigma_i(t)\sigma_i(t')-C_N(t,t'))^2>}
\label{fluct}
\end{equation}

In the equilibrium case the
fluctuations have a finite, extensive and not self-averaging value (see
\cite{macyou:sk}):
\[
\lim_{N\rightarrow\infty}\lim_{t_w\rightarrow\infty}F_N(t_w,t_w+\tau)=q_2-q_1^2
\label{flucteq}
\]
where $q_n=\int{q^n P(q) dq}$ is the n-th momentum of the equilibrium order parameter P(q).

\begin{figure}[] 
\epsfxsize=12cm
\epsfbox{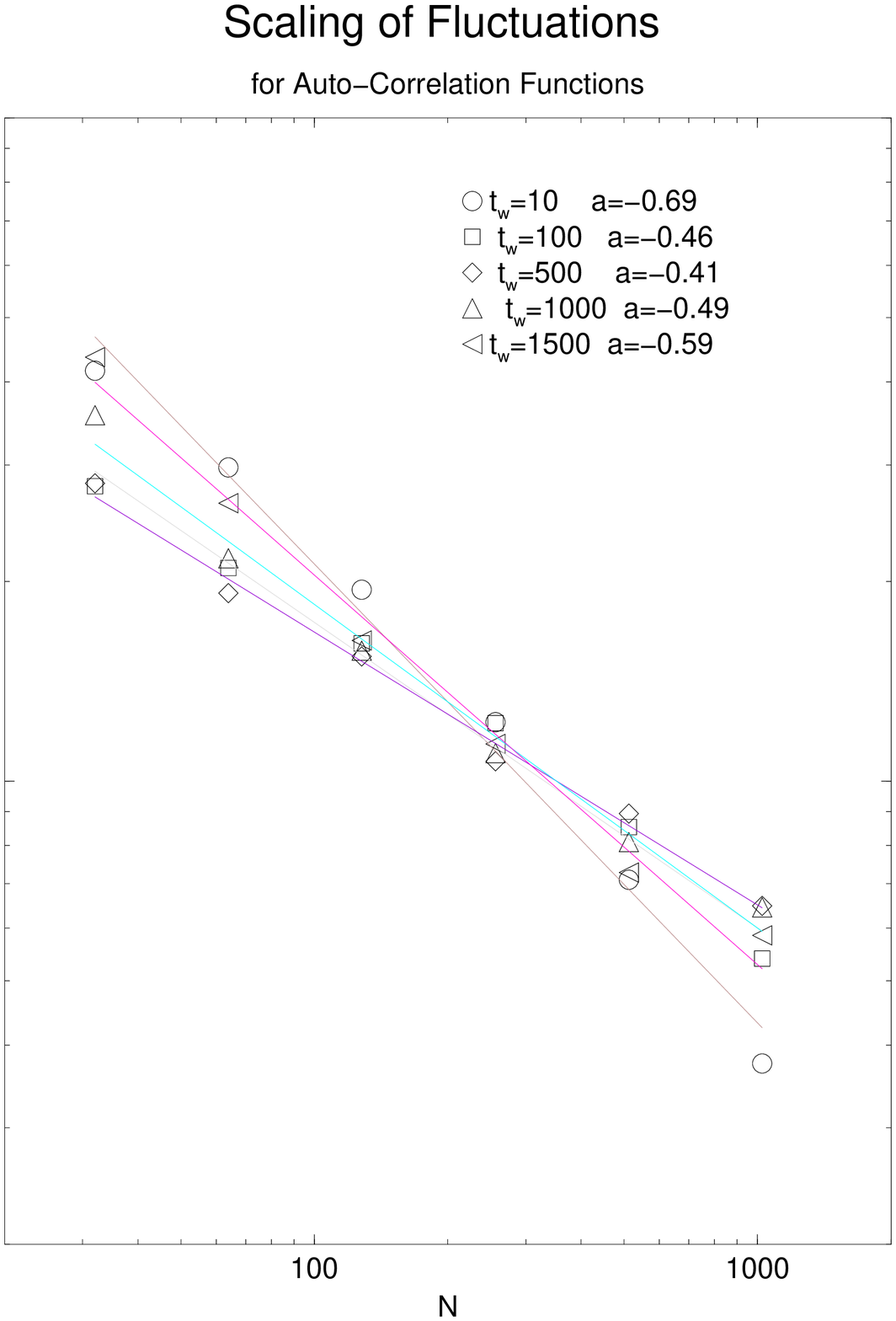}
\caption[]{Scaling of Auto-correlation Fluctuations
\label{figflut}}
\end{figure}

Instead, in the
{\em Out of Equilibrium Regime}
such quantities scale to zero with the size of the system.
To show this, we have integrated the fluctuations for several sized systems.
In Fig.\ref{figflut} we show the results for different 
values of $t_w$. We fit each curve with a power law $F_N \propto N^a$ (the result is shown in the legend).

\subsection{Clonation Procedures}

	To obtain a clearer picture of the geometrical
landscape, where the out of equilibrium dynamics take place,
we perform a particular simulation procedure that we call
``clonation'' \cite{bal:thesis,cugdea:clones,barburmez:clones}.
Since we are interested in the phace space
region visited by the dynamics after a certain time $t_w$,
we proceed to let evolve a single system with the usual
Monte Carlo dynamics, until such a time. At $t_w$
we create a number of copy of this system, i.e. systems exactly in the
same configuration (we {\em clone} it). 
Subsequently we let them (the original and the {\em clones}) evolve
independently, that is to say with the same Hamiltonian, but
different noise realizations. 

These systems represent different possible {\em histories} of
the first system, as if it had met, in its Monte Carlo
dynamics, a different random number sequence after
the time
$t_w$. There is a difference between these copies ({\em clones})
of the system and the previously defined {\em replicas}.
The {\em replicas} start at time
$t=0$ from independently choosen configurations and evolve
with independent dynamics (same Hamiltonian, different noises).
We may think at the {\em clones} like
starting from the same configuration at the time
$t=0$ and evolving identically (same Hamiltonian and same noise) 
until the time
$t=t_w$ (obviously this process is not actually simulated).

	This {\em clonation} procedure 
is different from a damage spreading procedure (see for example
\cite{der:damage, camdea:damage}), where two systems starting at measured
distance evolve whith the same noise. The damage spreading procedure was
used to individuate temperatures which could be compared with
equilibrium and dynamical transition temperatures.

	Our aim here is to investigate the geometry of the phase space
monitoring the Auto-Correlation Function (\ref{auto-correlaz}),
 and the {\em Clones-Correlation
Function}, defined by:

\begin{equation}
Q_{(N,t_w)}(t)=
\overline{<1/N\sum_{i=1,N}(\sigma_i(t)\tau_i(t))>}
\label{clones-correlaz}
\end{equation}
($\sigma$ e $\tau$ are the spins of two different clones).

These quantities are simply related to the
Euclidean distance in phase space.
	The Auto-Correlation (\ref{auto-correlaz}) is related 
to the Euclidean
distance between the configuration of each {\em clone} at
the time $t_w$ with its own configuration at the time
$t>t_w$ as:
\[
\overline{<\frac{1}{N}\sum_{i=1,N}(\sigma_i(t_w)-\sigma_i(t))^2>}=
2(1-C_N(t_w,t))
\]

Similarly the Clones-Correlation (\ref{clones-correlaz}) is related to the
distance between the {\em clones} (generated at the time $t_w$)
at the time $t$:
\[
\overline{<\frac{1}{N}\sum_{i=1,N}(\sigma_i(t)-\tau_i(t))^2>}=
2(1-Q_{(N,t_w)}(t))
\]

\begin{figure}[] 
\epsfxsize=12cm
\epsfbox{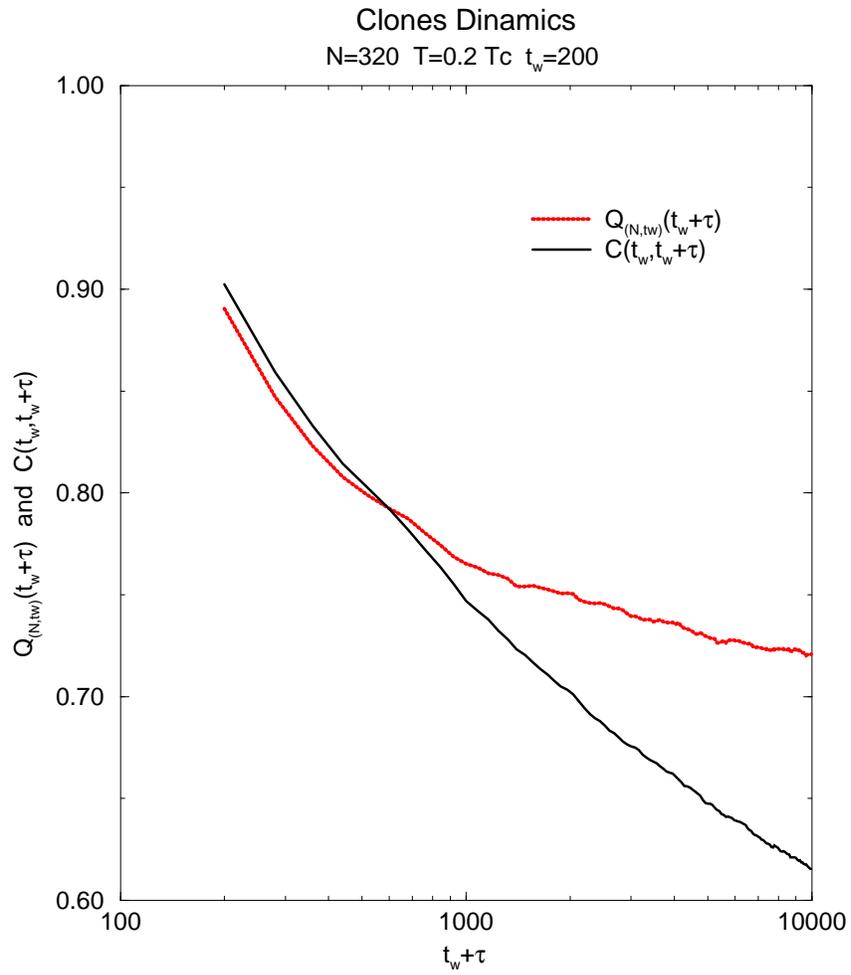}
\caption[]{Clonation of the system at the time $t_w$. the
plot shows the Auto-Correlation (\ref{auto-correlaz}) and
the Clones-Correlation (\ref{clones-correlaz}).
\label{figcloni}}
\end{figure}

In Fig.\ref{figcloni}
we show the results of the measurement.
We see that at first we have $C>Q$, but, after some time,
this relation inverts, and we have $Q>C$. 

Remembering the
relations of these quantities with the euclidean distances,
it means that at first the clones go away from each others
more than how much they drift away from the initial
configuration at the time $t_w$. But after, they continue
their drift standing close and forgetting the initial $t_w$
configuration. The simplest picture representing such a
behaviour is that of a dynamics following
canyons or in corridors: at first the clones span the width of
the channel and, after, they drift away along it. 

If the
situation has been that of a series of independent traps
(like for example in \cite{bou:aging}) we should see a different
behaviour, probably with $Q=C$. Here we see, at least, a kind
of hierarchy of traps \cite{boudea:aging}.

In the present work we do not get the asymptotic
limit of $Q$. We suspect that it goes to zero,
even if the curve is very slow but we can not exclude that
it may reach a costant value different from zero.

\section{Conclusions}
The non-equilibrium dynamics of the SK model takes place in
a range of time that grows exponentially with the size of
the system. In this region, we identify two distinct
regimes, characterized by different scaling properties of
$q_2(N,t)$ the average squared overlap beetween two typical
configurations visited at time $t$. The quantity
$q_2(N,t)$, for a system in equilibrium, represents the
second momentum of the
$P(q)$ distribution, the order parameter for the phase
transition.
In the case of a large system that starts the dynamics
from a random configuration, or that equivalently is
abruptely cooled from high temperature to a subcritical
one, the relevant regime is the first {\em Out of Equilibrium
Regime}, during which the $q_2(N,t)$ scales to zero as
$1/N$. In such a situation the dynamics is non-stationary,
presenting generic aging properties and confirming a
scenario of weak ergodicity breaking.  As claimed by recent
analitical works \cite{cugkur:pspin,cugkur:sk,balcugkurpar:sk},
 the asymptotic
configurations reached by an infinite SK model, present some
similarities with the equilibrium distribution: the out of
equilibrium staggered magnetisation equals the
equilibrium one and so does the energy density \cite{balcugkurpar:sk}.  But the configurations visited are not real
equilibrium configurations and the system allways escapes from them,
never to return.  These configurations  present a sort of
hierarchical structure. A special clonation procedure shows
that the dynamics takes place following corridors or canyons and
that phace space looks like a kind of fast flat labyrinth
that the system explores always slower looking for the
equilibrium configurations. 
Such a simple numerical procedure reveals that
the spin-glass dynamics is not simply a two-well energy problem, but we
have to consider more complex situations (see for example \cite{kurlal:surf}
or \cite{rit:backgammon})

\section*{Acknowledgments}
We thank L.F.Cugliandolo and J.Kurchan for their collaboration, 
and G.Parisi for the helpful supervision, troughout
the developing of this work. 



\end{document}